\documentclass[9pt,twocolumn,twosides]{revtex4}%
\usepackage{graphicx,amssymb,amsmath,color,psfrag}
\usepackage{amsthm}
\usepackage{amsfonts}
\usepackage{algorithmic}
\usepackage{enumerate}
\usepackage{latexsym}
\usepackage{amsmath}
\usepackage{amssymb}
\usepackage{textcomp}
\usepackage{graphicx}%
\usepackage{textcomp}
\usepackage[utf8]{inputenc}
\providecommand{\U}[1]{\protect\rule{.1in}{.1in}}
\begin{document}
\title{Light chaotic dynamics in the transformation from curved to flat surfaces}
\author{Chenni Xu$^{1,2}$}
\author{Itzhack Dana$^{1}$}
\author{Li-Gang Wang$^{2}$}
\author{Patrick Sebbah$^{1}$}
\thanks{patrick.sebbah@biu.ac.il}
\affiliation{$^{1}$ Department of Physics, The Jack and Pearl Resnick Institute for
Advanced Technology, Bar-Ilan University, Ramat-Gan 5290002, Israel}
\affiliation{$^{2}$ Zhejiang Province Key Laboratory of Quantum Technology and Device,
Department of Physics, Zhejiang University, Hangzhou, 310027, Zhejiang, China}

\begin{abstract}
Light propagation on a two-dimensional curved surface embedded in a three-dimensional space has attracted increasing attention as an analog model of four-dimensional curved spacetime in laboratory. Despite recent developments in modern cosmology on the dynamics and evolution of the universe, investigation of nonlinear dynamics of light in non-Euclidean geometry is still scarce and remains challenging. Here, we study classical and wave chaotic dynamics on a family of surfaces of revolution by considering its equivalent conformally transformed flat billiard, with nonuniform distribution of refractive index. This equivalence is established by showing how these two systems have the same equations and the same dynamics. By exploring the Poincar\'{e} surface of section, the Lyapunov exponent and the statistics of eigenmodes and eigenfrequency spectrum in the transformed inhomogeneous table billiard, we find that the degree of chaos is fully controlled by a single geometric parameter of the curved surface. A simple interpretation of our findings in transformed billiards, the "fictitious force", allows to extend our prediction to other class of curved surfaces. This powerful analogy between two a prior unrelated systems not only brings forward a novel approach to control the degree of chaos, but also provides potentialities for further studies and applications in various fields, such as billiards design, optical fibers, or laser microcavities.

\end{abstract}
\maketitle




\section{Introduction}

Einstein's general theory of relativity (GR) unprecedentedly interprets
gravity in a geometrical framework, namely, the presence of a massive object
distorts the very fabric of the space and time in its vicinity.
Despite of GR's triumph, 
gravitational effects are in principle too feeble for light to be perceived in laboratory environment. 
One of the analog models of GR is two-dimensional (2D) curved surfaces embedded in 3D space, whose theoretical cornerstones are the 3+1 membrane paradigm \cite{Thorne1986} and embedding diagrams. After taking a constant time and extracting the equatorial slice of a spherically symmetrical space, the remnant metric can be visualized through a 2D curved surface.
Ever since this innovative notion was brought up by Batz and Peschel \cite{Batz2008}, electromagnetic (EM) dynamics and wave propagation on
2D curved surfaces have been prosperously developed both theoretically
\cite{Batz2010, Bekenstein2014, Lustig2017, XuPRA, XuOE, XuNJP, Wang2018,
Jing2021} and experimentally \cite{Schultheiss2010, Schultheiss2015,
Patsyk2018, Bekenstein2017, Li2018}. 
On the other hand, GR notions could reciprocally contribute to the engineering of nanophotonic devices \cite{Bekenstein2017, Lebental2021} and transformation optical designs, ranging from EM invisible cloaks \cite{Tyc2009, Zolla2009} to perfectly focusing lens \cite{LXu2019, LXu2019OE, Patrick2015}, based on manipulation of the curved geodesic paths.

Investigation of nonlinear dynamics in the context of GR may date back to 1960s, and has been extensively discussed in various aspects \cite{Radu2017, Contopoulos2008, Aydiner2018}. Nonlinear dynamics, or chaos, widely exists in diverse modern scientific
disciplines
. One characteristic of chaos is the exponential sensitivity to initial conditions, 
with\emph{ }a well-publicized metaphor being
the \textquotedblleft butterfly effect\textquotedblright. In the recent decades, instead of being considered a nuisance, wave chaos has been taken advantage of to ameliorate optical resonators and laser microcavities \cite{Liu2013, Cao2018, Cho1998, Stone1997, Kim2004, Hentschel2008, Xiao2017, Rotenberg2015, Fratalocchi2014,  
Fan2021}, such as enhancing energy storage \cite{Liu2013}, suppressing spatiotemporal lasing instabilities \cite{Cao2018}, realizing high power directional emission \cite{Cho1998, Stone1997, Kim2004}, etc. To this end, information about dynamical behaviors in 2D cavity, such as critical periodic orbits and volume of chaotic area in phase space, is significant. One paradigm model is a 2D table (or flat) billiard, 
which is typically a closed domain wherein particles and light rays propagate freely, except for elastic collisions on its boundaries. Various degrees of chaos in table billiards are realized by either deforming the boundaries \cite{Sinai1963, Berry1981, Bunimovich2001} or introducing external force \cite{Leonel2015, Berry1985, Goodings1993}. In a recent work, a novel notion, transformed cavity, is proposed with a nonuniform profile of refractive index present in a deformed optical cavity \cite{Min2016}. Such transformed cavity opens up a new pathway of chaos engineering.

Investigation of light chaotic dynamics in a non-Euclidean geometry remains challenging from the complexity of the equations induced by the curved geometry, as well as the ambiguity of concepts, such as the definition of boundaries and indicators of chaos.
To the best of our knowledge, the wave chaos idea has been carried over in a few curved elastic systems \cite{Sodergaard2010, Sodergaard2007, Rebinsky1994}, such as aluminum thin shells \cite{Sodergaard2010}, yet has rarely been explored in optics, except for a very recent study about ray chaotic behavior on a deformed toroidal surface \cite{Chan2020}. Here we address this challenge by developing the analogy between a curved surface and a 2D table billiard with nonuniform distribution of refractive index, on the strength of transformation optics (TO). 
The mathematical underpinning of TO is the form invariance of Maxwell equations under general coordinate transformation, with the optical constitutive parameters of the transformed media (i.e., engineering the permittivity $\varepsilon$ and permeability $\mu$ tensor explicitly) \cite{Pendry2006, Schurig2006}. As we will demonstrate, this approach of projection provides a possibility to utilize physical systems on one surface to explore their counterparts on the other, especially when the studies in one of the systems are demanding to carry on.

In this work, we study classical and wave chaos on a special type of 2D curved surfaces in 3D space, surfaces of revolution (SORs), by investigating light propagation in its corresponding table billiard with azimuthally symmetric nonuniform distribution of refractive index. We exemplarily choose a typical family of SORs, the Tannery's pears. Since such surfaces are integrable, we consider half of the pear and its 2D nonuniform billiard where a circular hole is pierced to introduce chaos. We first establish that these two systems share the same dynamics by showing that a conformal coordinate transformation preserves both geodesic equations and wave equations. Thanks to this equivalence, we can assess the degree of chaos in one system (the curved surface) by probing the nature of the trajectories in Poincar\'{e} surface of section (SOS) and measuring the Lyapunov exponent in the other (the flat billiard). We find that the degree of chaos is fully controlled by a single geometrical parameter of the Tannery's pear. This is also revealed in the statistics of eigenmodes of the Tannery's pears, as the wave equation is invariant for both systems. At last, we show that the above results can be generalized to arbitrary SORs, by coming up with a universal quantity in the 2D nonuniform billiards.\ This simple interpretation further demonstrates the potentialities of our approach.\

\section{Results}
\subsection*{Basic theory}

As its name indicates, a SOR can be generated by revolving an arbitrary plane
curve (known as generatrix) around an axis of symmetry for a circle. 
The line element of a typical
SOR takes the form $ds^{2}=g_{\alpha\beta}dx^{\alpha}dx^{\beta}=E(u)du^{2}+G(u)dv^{2}$,
with $u$ along longitudinal direction and $v$, being the rotational angle,
along transverse direction, as shown in Fig. 1A.
In curved spaces, light rays propagate along geodesics, in analogy with straight lines in free space. On the basis of geodesic equations, an arbitrary geodesic with given initial conditions is depicted by (for mathematical details, see SI Appendix, section 1)
\begin{equation}
dv=\eta\frac{\kappa\sqrt{E(u)}}{G(u)\sqrt{1-\frac{\kappa^{2}}{G(u)}}}du.
\label{uv}%
\end{equation}
Here $\eta=\text{sgn}\left[\left(du/ds\right)_{\text{initial}}\right]$ is determined by the initial direction, and slant $\kappa$, defined as $\left[G(u)dv/ds\right]_{\text{initial}}$, remains invariant as a consequence of the conservation of angular momentum \cite{Wang2018}. When it comes to waves, the time-harmonic scalar EM field follows the massless Klein-Gordon equation
\begin{equation}
\frac{1}{\sqrt{E(u)G(u)}}\frac{\partial}{\partial u}\left[  \sqrt{\frac
{G(u)}{E(u)}}\frac{\partial\Phi}{\partial u}\right]  +\frac{1}{G(u)}%
\frac{\partial^{2}\Phi}{\partial v^{2}}+k^{2}\Phi=0, \label{we11}%
\end{equation}
with $k$ being the propagation constant \cite{note}.

In this work, instead of directly investigating light propagation on the surface, we employ an innovative method by projecting a SOR onto a plane with distribution of refractive index, in light of conformal transformation optics \cite{Leonhardt2006, Chen2015}. This notion is mathematically underlain by the theorem that any 2D Riemannian manifold is conformally flat. Namely, the metric of an arbitrary 2D curved surface $ds^2$ can be conformally related to the metric of a flat plane $ds^2_\text{f}$ through $ds^2=\Lambda ds^2_\text{f}$, where $\Lambda$ is a differentiable function \cite{inverno1992}. Interestingly, the right-hand side, defined as $ds^{\prime2}$, perfectly describes an inhomogeneous planar dielectric medium with spatially varying refractive index $n^{\prime}=\sqrt{\Lambda}$. For SORs, thanks to their rotational symmetry, we can naturally suppose the projected plane is azimuthally symmetric. Consequently, polar coordinate is chosen out of convenience and the variation of refractive index rests exclusively on radial component. Based on the premise that $ds^2=ds^{\prime2}$, an equivalence between these two systems can be established through the coordinate mapping
\begin{equation}
r(u)=A\exp\left[  \int^{u}\sqrt{\frac{E(u^{\prime})}{G(u^{\prime})}}%
du^{\prime}\right]\text{,}\ v=\varphi\text{,} \label{rrr}%
\end{equation}
with distribution of refractive index
\begin{equation}
n^{\prime} (u)=\frac{\sqrt{G(u)}}{A}\exp\left[  -\int^{u}\sqrt{\frac
{E(u^{\prime})}{G(u^{\prime})}}du^{\prime}\right]  . \label{nnn}%
\end{equation}
Here $A$ is a positive integration constant, $r$ and $\varphi$ are radial and azimuthal coordinates, respectively.

By virtue of this coordinate mapping, transformed version of Eq. (\ref{uv}) appears as
\begin{equation}
d\varphi =\eta ^{\prime }\frac{\kappa ^{\prime }}{n^{\prime }(r)r^{2}\sqrt{1-%
\frac{\kappa ^{\prime 2}}{n^{\prime 2}(r)r^{2}}}}dr,
\label{geodesic2222}
\end{equation}
with $\eta ^{\prime }=\text{sgn}\left[ \left( dr/ds^{\prime }\right) _{\text{%
initial}}\right] $ and $\kappa ^{\prime }=\left[ n^{\prime 2}(r)r^{2}d\varphi
/ds^{\prime }\right] _{\text{initial}}$. We prove in SI Appendix 2 that Eq. (\ref{geodesic2222}) is the very solution of geodesic equations on the projected plane. Thus the dynamics of light rays are preserved in both systems. In terms of waves, Eq. (\ref{we11}) is transformed into%
\begin{equation}
\frac{1}{r}\frac{\partial}{\partial r}\left(  r\frac{\partial\Phi}{\partial
r}\right)  +\frac{1}{r^{2}}\frac{\partial^{2}\Phi}{\partial\varphi^{2}}%
+k^{2}n^{\prime2}(r)\Phi=0, \label{we22}%
\end{equation}
which is exactly the wave equation of electric field on a plane with distribution of azimuthally symmetric refractive index $n^{\prime}(r)$. As a result, the transformation preserves the solutions of wave equation in one system to be the eigenmodes of another. In SI Appendix 2, we extend this result to arbitrary curved surfaces, demonstrating the universality of the transformability of both geodesic equations and wave equations. We have shown, therefore, that the dynamical properties of both optical rays and waves are absolutely identical on a homogeneous SOR and on its transformed inhomogeneous plane, and we can safely investigate one system to infer about the other.

\begin{figure}[htb]
\centering
\includegraphics[width=6cm]{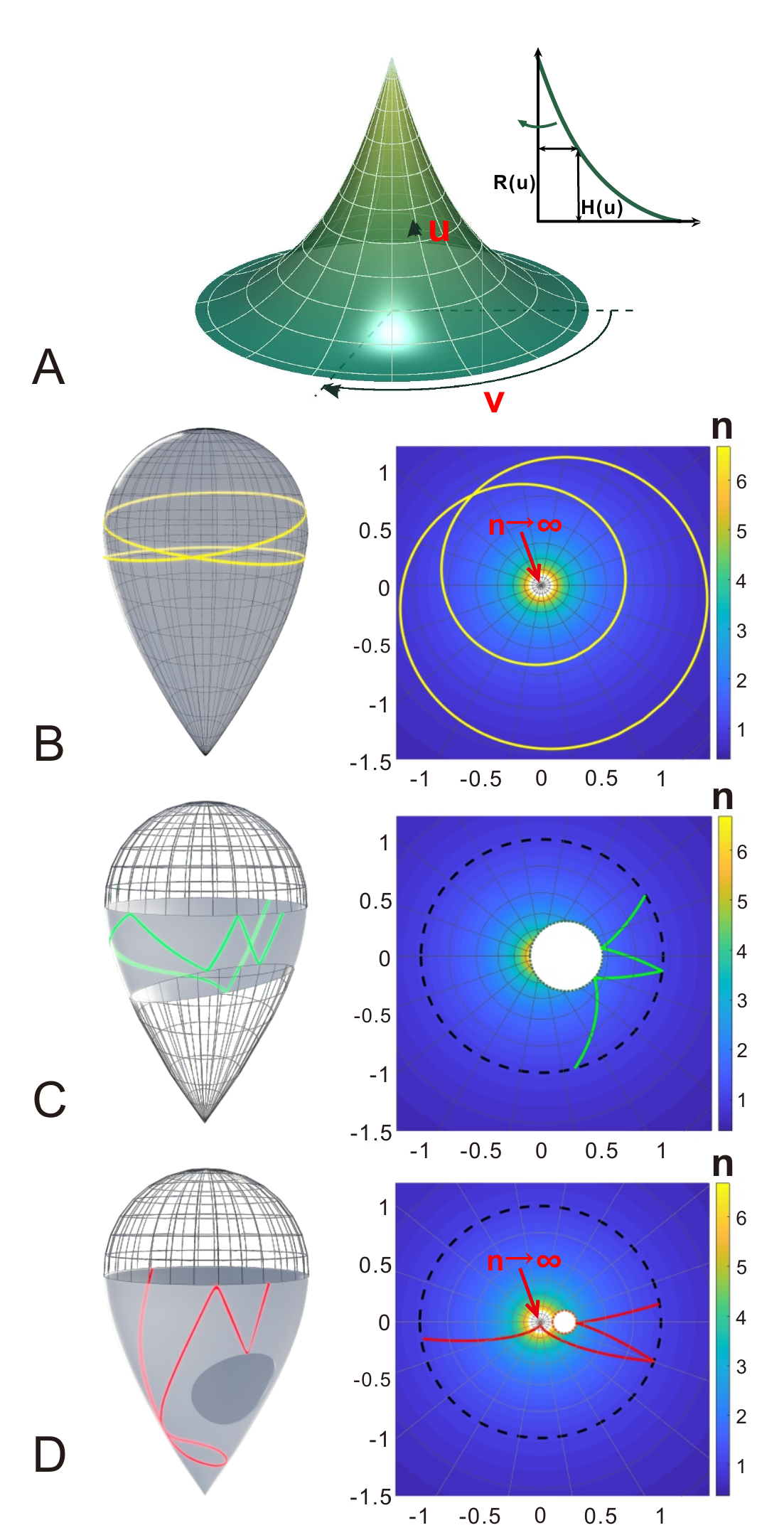}\caption{(A) Sketch of a typical SOR with the orthogonal curvilinear coordinates on it. The inset illustrates its generatrix. (B) Tannery's pear with $c=2$ (left) and its projected billiard (right). A typical closed trajectory is depicted by yellow solid line.\ (C),(D) Truncated Tannery's pears which have a circular hole with (C) radius $0.3$ (dark green dotted line) and (D) radius $0.1$ (orange dotted line) in their projected billiards. The equators are denoted by black dashed
lines in the projected billiards. Two typical trajectories are plotted by green and red solid lines, respectively.}%
\label{figure1}%
\end{figure}

When a SOR is closed (e.g. sphere, spindle) or infinitely extends (e.g.
cylinder, cone), its projection covers the whole plane. A paradigmatic example
is the so-called \textquotedblleft Tannery's pear\textquotedblright, which is
an object of interest for mathematicians \cite{Besse1978} and is going to be
the study case of this paper. The family of Tannery's pears can be
parametrized by $E(u)=(c+\cos u)^{2}$, $G(u)=\sin^{2}u$, with $c$ being a
positive constant and in the meantime, the single parameter to describe this
family of surfaces. It is proved that such surfaces have a constant period
function $2c\pi$ \cite{Borzellino2006}. That is, when the parameter $c$ is an
integer, all the geodesics (except the ones along longitudinal $u$ direction)
are closed, and for a particle which departs from an arbitrary position on the
geodesic, moves along it and returns to the starting point, the accumulated
variation in coordinate $v$ is $2c\pi$, as shown in Fig. 1B. In order to limit our study to a
finite area on plane, a mirror is put on the \textquotedblleft
equator\textquotedblright\thinspace\ (i.e., the latitude with $u=\pi/2$, see
the black dashed lines in Figs. \ref{figure1}C and \ref{figure1}D) of the
surface, which plays the role of a perfectly reflecting boundary, and only the
lower half of surface is taken. On the projected plane, the equator
corresponds to a circle with unit radius, while the lower half of surface
corresponds to the area inside the circle, which together naturally form a
circular billiard, as shown in the right column of Figs. \ref{figure1}C
and \ref{figure1}D. These two systems are absolutely integrable, since the
number of constants of motion (energy and orbital angular momentum) equals to
the dimensionality of the system \cite{Gutzwiller1990}, owing to the
rotational symmetry of SORs. To introduce chaos in this system, an
off-centered disk-shaped area is eliminated, leaving a circular hole with a
specular boundary in the billiard. The choice of a disk shape is out of
consideration of simpleness, yet more complex shapes could be explored. The
right column of Figs. \ref{figure1}C and \ref{figure1}D shows two typical
holes whose centers are located at $0.2$ (in unit of radius) from the centers
of the billiards, and whose radius are $0.3$ and $0.1$, denoted by dark green
and orange dotted lines, respectively. By taking the coordinate transformation Eq.
(\ref{rrr}), the correspondence of these two holes on the surface are shown in
the left column. It is seen that the refractive index approaches to
infinity at the center of the projected billiard, corresponding to the apex of
the surface in the bottom, which is basically a singularity for light rays. We
focus in the rest of the paper on the cases where the hole includes the
billiard center, resulting in the removal of the bottom apex. Its
correspondence on the surface is a boundary unparallel to latitudes, rather
than a hole.

In what follows, in order to distinguish them from their 2D projected
billiards, SORs are referred to as \textquotedblleft3D
surfaces\textquotedblright, based on the fact that they are embedded in 3D
space in spite of being 2D per se.

\subsection*{Poincar\'{e} surface of section}

\begin{figure*}[htb]
\centering
\includegraphics[width=14.5cm]{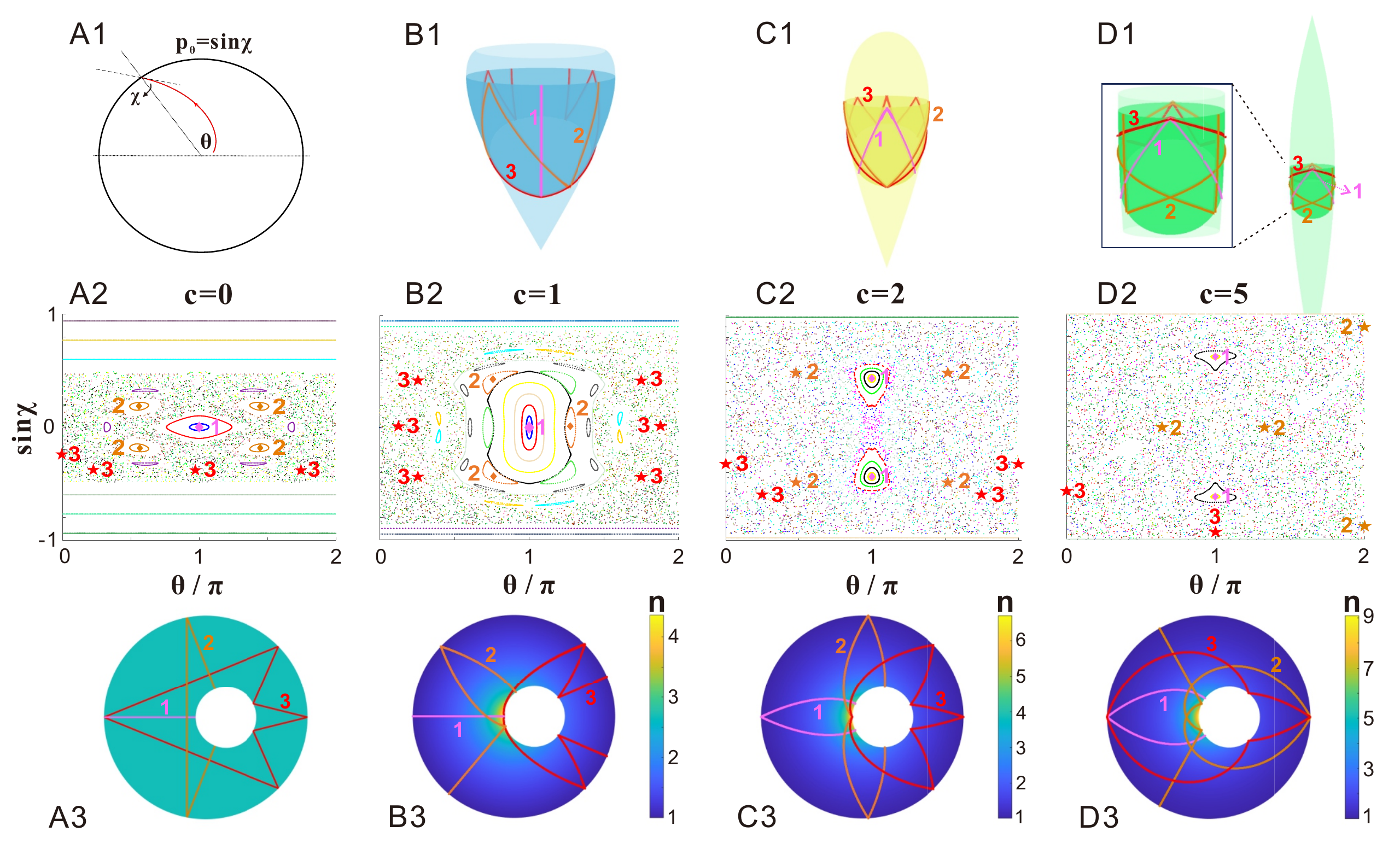}\caption{Poincar\'{e} surfaces of
section of Tannery's pears with different parameter $c$. (A1) Sketch of
Birkhoff coordinates. (B1),(C1),(D1) Sketches of Tannery's pears with $c=1$,
$2$ and $5$, respectively. The transparent parts correspond to areas outside
the projected billiard and inside the hole, which are excluded in the study.
(A2),(B2),(C2),(D2) Poincar\'{e} surface of section of Tannery's pears with
$c=0$, $1$, $2$ and $5$, respectively. Each color represents an arbitrary
trajectory. The diamonds and pentagrams respectively indicate typical stable
and unstable periodic orbits, whose representations in Poincar\'{e} surface of
section are a succession of points. These periodic orbits in real space are
exhibited in (A3), (B3), (C3), (D3), respectively, corresponding to the points
with the same label and color.}%
\label{figure2}%
\end{figure*}

Ray trajectories become rapidly intractable in real space after few bounces, it is therefore more advisable to inspect ray motions in phase space which are composed of two spatial dimensions and their conjugate momenta. For billiard systems, an effective approach to extract information from 4D phase space is to record the states of the trajectory exclusively when it collides on the outer boundary. This 2D section of phase space is known as Poincar\'{e} surface of section (SOS), and is conveniently depicted in the framework of Birkhoff coordinate, as sketched in Fig\textbf{.} \ref{figure2}A1.
In practice, the trajectories between two bounces are traced by Eq. (\ref{uv}). When the trajectories collide on the hole/lower
unparallel boundary, the rule of specular reflection is applied in the 2D projected billiard, after projecting the trajectories on 3D surface back to the inhomogeneous plane. This step saves one from complicated or even impractible calculation on 3D surface, since the analytical expression of the lower unparallel boundary is inaccessible.

\begin{figure}[htb]
\centering
\includegraphics[width=7.5cm]{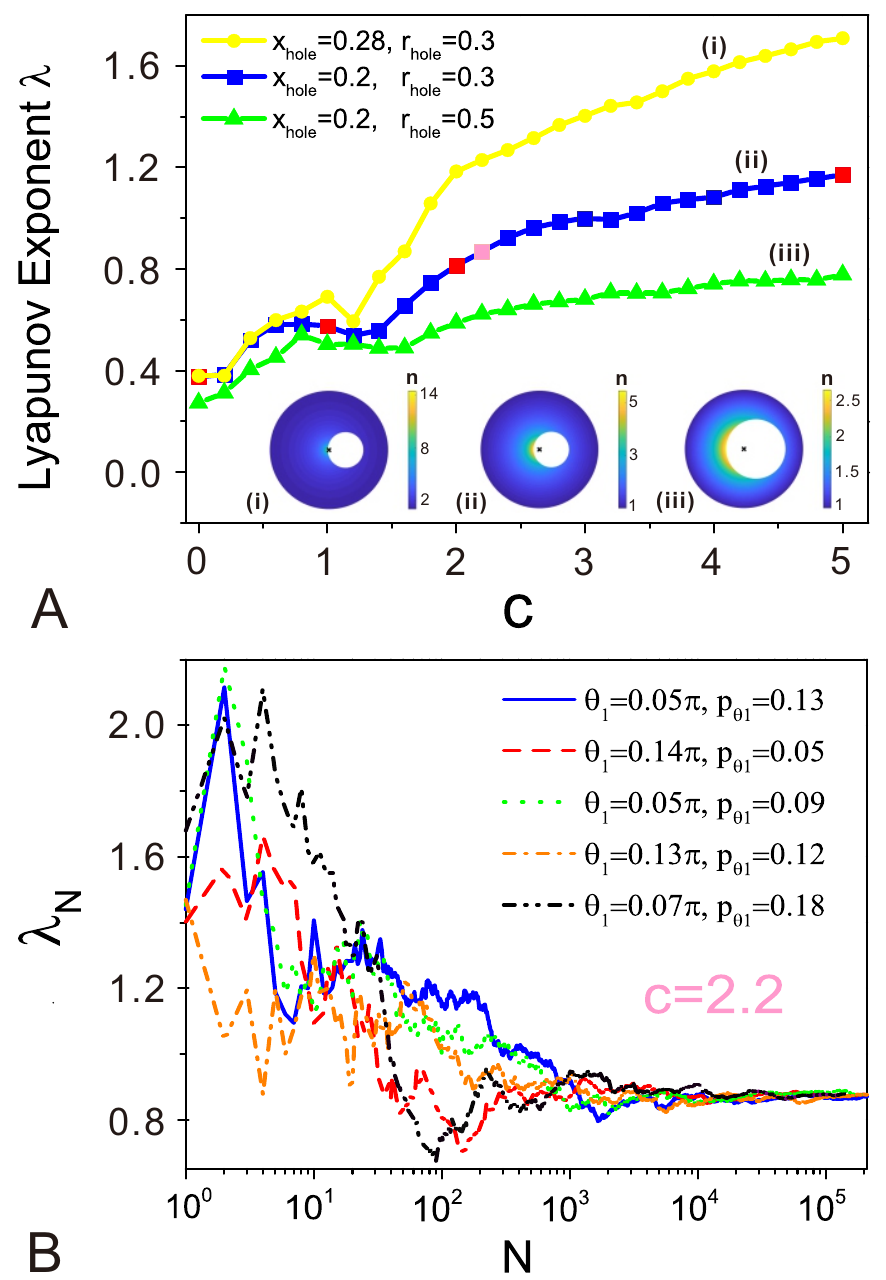}\caption{(A) Variation of Lyapunov
exponents versus the parameter $c$ of Tannery's pears and three different means of truncation.
The distribution of refractive index are illustrated in the inset with corresponding labels, and the black cross denotes the center of the billiard. The red square data markers in
line (ii) correspond to the cases illustrated in Fig. \ref{figure2}. 
(B) The quantity $\lambda_{N}$ at $N$th bounce, calculated by Eq. (\ref{le2}).
Five pairs of extremely close trajectories are randomly chosen at the case denoted by the pink
square data marker in (A), whose initial conditions are shown in the legend with $\theta_{2}=\theta_{1}+10^{-4}$, $p_{\theta_{2}}=p_{\theta_{1}}+10^{-4}$. 
All the five pairs of trajectories converge to the same asymptotic value,
which is defined as Lyapunov exponent. }%
\label{figure3}%
\end{figure}

When $\left\vert \sin\chi\right\vert $\ is large enough, trajectories
propagate near the outer boundary of 2D\ projected billiard/equator of
3D\ surface and are therefore free from colliding on the hole/lower unparallel
boundary. Their tangential momenta $\left\vert \sin\chi\right\vert $\ remain
invariant, so that in Poincar\'{e} SOS these trajectories are represented by
horizontal straight lines. 
In contrast, trajectories with small $\left\vert \sin\chi\right\vert $ collide on the
hole/lower unparallel boundary, resulting in their representations ergodically
and irregularly distributed in the central part of Poincar\'{e} SOS, which
therefore forms a chaotic sea. 
Among these trajectories, some return to their initial conditions after several bounces, and retrace themselves repeatedly. Such periodic orbits (POs) are
represented as a succession of points in SOS, with the number of points equal to the number of bounces on the outer boundary/equator in each period. For
some POs, the trajectories whose initial conditions are slightly deviated from the critical conditions of POs are able to survive in the vicinity of the latter. Such POs, considered as stable POs, form regular islands among the chaotic sea in Poincar\'{e} SOS together with their quasi-periodic orbits. As a contrast, the other unstable POs are submerged in the chaotic sea and can be scarcely recognized. Some typical stable and unstable POs are illustrated by diamonds and pentagrams respectively in the middle row of Fig. \ref{figure2}, and the corresponding trajectories in the billiards are depicted in the last row with same labels and colors. An interesting feature of PO with label $1$ is explicated in SI index, section 3.

A generic presentation about the Poincar\'{e} SOSs of four
typical Tannery's pears with $c=0$, $1$, $2$, $5$ is given in Fig.
\ref{figure2}. Different shapes of the surfaces are clearly exhibited in Figs.
\ref{figure2}B1, \ref{figure2}C1 and \ref{figure2}D1. Specially, when
$c=0$, the metric of the surface degenerates to $ds^{2}=\cos^{2}udu^{2}%
+\sin^{2}udv^{2}$, which is equivalent to a polar coordinate if one takes
$\sin u$ as the radial component and $v$ as the azimuthal component. Thus,
surface with $c=0$ corresponds to the flat billiard and serves as reference. The Poincar\'{e} SOSs reveal that the phase spaces of all the four
surfaces are mixed, with coexistence of both regular and chaotic trajectories.
One can observe that the area of chaotic sea remarkably enlarges with the
parameter $c$, implying an increasingly chaotic phase space structure of
Tannery's pear with greater parameter $c$. Another phenomenon one may
meanwhile notice is an obvious reduce in the amount of stable POs and the area
of islands, especially in Figs. \ref{figure2}B2, \ref{figure2}C2 and
\ref{figure2}D2.\emph{ }These two hints suggest an increasing proportion of
trajectories transferring from regular to chaotic, and consequently indicate a
more significant degree of chaos.\emph{ }Note that\emph{ }this observation is
true both on 3D surface and in 2D billiard with refractive index landscape, because of their equivalence.\emph{ }

\subsection*{Lyapunov exponent}

As was mentioned,  
a hallmark of chaos is the sensitivity to initial conditions, i.e., the exponential divergence of two extremely nearby trajectories in phase space. The speed of this exponential divergence is characterized by the (maximum) Lyapunov exponent $\lambda$. Here, we
adopt the method of \textquotedblleft billiard map\textquotedblright%
\thinspace\ in terms of the collisions of trajectories on the outer boundary, instead of the \textquotedblleft billiard flow\textquotedblright%
\thinspace\ with continuous time \cite{Datseris2019}. In this way, Lyapunov exponent is defined as
\begin{equation}
\lambda=\lim_{N\rightarrow\infty}\frac{1}{N}\sum\limits_{i=1}^{N}\ln
\frac{\left\vert \boldsymbol{\delta}_{i}\right\vert }{\left\vert
\boldsymbol{\delta}_{0}\right\vert }, \label{le2}%
\end{equation}
where $\left\vert\boldsymbol{\delta}_i\right\vert $ is the distance of the two chosen trajectories in phase space at the $i$th bounce. In practice, two sets of extremely close initial conditions are randomly chosen from the chaotic area in Poincar\'{e} SOS, and distance $\left\vert\boldsymbol{\delta}\right\vert $ is collected everytime two trajectories collide on the outer boundary. Theoretically, if the number of bounces $N$ is large enough, Eq. (\ref{le2})
is supposed to approach to a constant which is independent of the choice of initial conditions.  
Technical details about the calculation of Lyapunov exponents are specified in SI index, section 4.

By this method, Lyapunov exponents of a series of Tannery's pears with varying
parameter $c$\ and under three different means of truncation are calculated
and illustrated in Fig. \ref{figure3}A.
Each data point in Fig. \ref{figure3}A is the result of averaging over 4-6
randomly-chosen pairs of trajectories, where each pair of trajectories
experiences several tens of thousands of bounces on the outer
boundary/equator, and an approximate convergence has been reached (Example of
five typical pairs of trajectories is shown in Fig. \ref{figure3}B). The
figure manifests an overall increasing tendency of Lyapunov exponents with
increasing parameter $c$\ of Tannery's pears, indicating that in general, when
parameter $c$ increases, the speed of divergence between two close
trajectories, or the instability of the trajectories increases, which
validates the increase of degree of chaos both on the 3D surface and in the 2D
projected billiard. Furthermore, the invariance of this increasing tendency in
spite of different positions and radii of the pierced hole in the projected
billiard attests the universality of this feature, as shown in Fig.
\ref{figure3}A. Compared with the qualitative interpretation obtained from
Poincar\'{e} SOS with a coarse sampling of parameter $c$, this signature
brings a more quantitative insight, which enables a finer exploration and
hence reveals more details about the dependence of chaoticity on parameter
$c$. One example is that an exception to this increasing trend is observed
near $c=1$, elucidating that the variation of chaoticity with parameter $c$ is
not strictly monotonous.

\subsection*{Statistics of eigenmodes and eigenfrequency spectrum}


\begin{figure*}[htb]
\centering
\includegraphics[width=12.5cm]{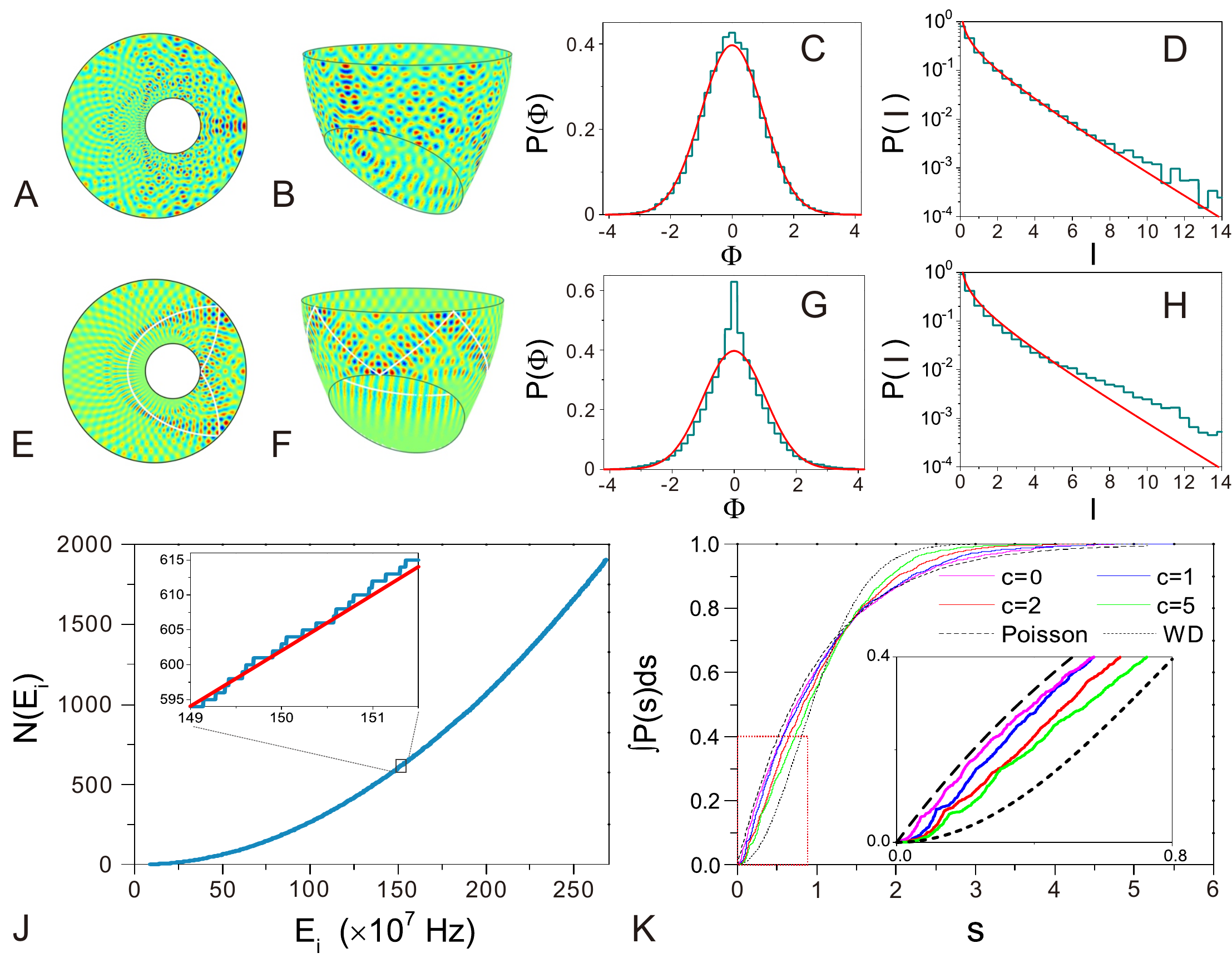}\caption{(A)-(H) Spatial statistics of eigenmodes. (A),(B) The field distributions (amplitude) of a typical ergodic eigenmode (frequency $\omega=2.9639$ GHz) of truncated Tannery's pear with $c=1$ in projected billiard (A) and on 3D\ surface (B). Its distribution of amplitude and intensity are plotted in (C) and (D), respectively, with red solid line being Gaussian and Porter-Thomas distribution as reference. The field distributions (amplitude) of a typical scarring eigenmode (frequency $\omega=2.8863$ GHz) and their statistics are shown in (E)-(H). The white
solid lines in (E) and (F) indicate its corresponding unstable periodic orbit. (J),(K) Statistics of eigenfrequency spectrum. (J) Cumulative eigenfrequency density $N(E_{i})$ of Tannery's pear with $c=2$
versus eigenfrequency $\left\{  E_{i}\right\}  $. (K) Distribution of nearest neighbor
spacing of Tannery's pears with $c=0$ (purple), $1$ (blue), $2$ (red) and $5$
(green). The Poisson (black dashed) and Wigner-Dyson (black short dashed)
distribution are also plotted as reference. (Inset) Zoom-in of the area marked
by red dotted rectangle. 
}%
\label{figure55}%
\end{figure*}

When it comes to quantum or wave realm, 
notions of phase space and trajectories are not properly defined due to the uncertainty principle.
Therefore, signatures in classical chaos are no longer valid, but are replaced by quantum mechanical criteria based on energy spectrum, energy eigenvectors, entanglement, temporal evolution of expectation values, etc. In this section, we are going to focus on the first two fingerprints.


We have demonstrated that the wave equation on a SOR  
is identical with that in its projected billiard,
bridged by the coordinate transformation. 
Namely, the performed calculation as well as the
concepts and phenomena in one system can be automatically extended to the other. Here, we obtain eigenmodes in both systems with the help of COMSOL Multiphysics 5.2. In practice, the simulation is conducted in the 2D inhomogeneous billiard, because of the difficulty in constructing 3D surface (especially when the expression of the lower unparallel boundary is unknown), and the eigenfunction on 3D surface is subsequently obtained by projection from 2D billiard via the coordinate transformation. Three typical eigenmodes in both systems are illustrated in SI index, Fig. S4.

One credible imprint of ray chaos resides in the spatial statistics of eigenmodes 
\cite{Legrand2013}. To be specific, the wave function of a typical ergodic eigenmode, whose classical correspondence has a stochastic motion, distributes uniformly over the
whole available area of the phase space which is ergodically visited by its
classical trajectories. 
Such modes are thus conjectured as
a random superposition of plane waves with different amplitudes, phases and
directions but same wavenumber, manifesting an analogous pattern of laser
speckles. 
As per central limit theorem, such
field is Gaussian random, 
implying the amplitude of
eigenmode follows the Gaussian statistical distribution
\begin{equation}
P\left(  \Phi\right)  =\frac{1}{\sqrt{2\pi\left\langle \Phi\right\rangle }%
}\exp\left(  -\frac{\Phi^{2}}{2\left\langle \Phi\right\rangle }\right)  ,
\label{Pfai}%
\end{equation}
while the probability distribution of intensity $I=\Phi^{2}$ is subjected to
\begin{equation}
P\left(  I\right)  =\frac{1}{\sqrt{2\pi I/\left\langle I\right\rangle }}%
\exp\left(  -\frac{I}{2\left\langle I\right\rangle }\right)  , \label{PI}%
\end{equation}
known as Porter-Thomas (PT) distribution. Figs. \ref{figure55}A and \ref{figure55}B
illustrate a typical ergodic eigenmode of truncated Tannery's pear with $c=1$,
both on 3D surface and 2D projected plane. The good agreement of its $P\left(
\Phi\right)  $ and $P\left(  I\right)  $ with Gaussian and PT distribution
respectively validates the ergodicity of the eigenmode.

Besides ergodic modes with speckle statistics, there exists a special class of modes with enhanced amplitude in the vicinity of single unstable periodic orbits in the corresponding classical system. This ubiquitous yet remarkable phenomenon is well known as Quantum Scarring \cite{Heller1984}. Unlike the enhancement of eigenstate intensity near stable periodic orbits which are well understood by
semiclassical theory of integrable systems,  
scar was initially an unexpected phenomenon and later found to be a significant correction to predictions from random matrix theory 
and Gutzwiller trace formula. 
One of scarred modes of truncated Tannery's
pear with $c=1$ on projected plane and 3D surface is exhibited in Figs.
\ref{figure55}E and \ref{figure55}F, superimposed by a white solid line indicating its corresponding unstable periodic orbit. A prominent deviation of $P\left(
\Phi\right)  $ and $P\left(  I\right)  $ from ergodicity can be clearly
observed in Figs. \ref{figure55}G and \ref{figure55}H respectively.

Another classic and widely recognized indicator for the randomness of a quantum system is the distribution of nearest neighbor spacing (NNS) of
eigenfrequency spectrum \cite{Zimmermann1986}.  
In a classically integrable system, successive eigenfrequency distributes randomly, and the NNS $s$ follows the Poisson law
\begin{equation}
P_{\text{Poisson}}(s)=\exp(-s), \label{Poisson}%
\end{equation}
with its peak located at zero. While in classically chaotic systems, the presence of level repulsion leads NNS to best fit Wigner-Dyson (WD) distribution%
\begin{equation}
P_{\text{WD}}(s)=\frac{\pi s}{2}\exp\left(  -\frac{\pi}{4}s^{2}\right)  .
\label{WD}%
\end{equation}
Contrast with Poisson distribution where level spacing can be zero with highest probability, a salient consequence of level repulsion is the
vanishment of infinitesimal NNS, i.e., $P(s)\rightarrow0$ when $s\rightarrow
0$. This behavior can be well explained by the Bohigas-Giannoni-Schmit conjecture 
\cite{Schmit1992} that spectra of time-reversal-invariant systems reveal the same fluctuation properties as a Gaussian orthogonal ensemble in random matrix theory. When the system has mixed dynamics, its distribution of NNS is intermediate between the two limiting cases. 
In virtue of COMSOL Multiphysics 5.2, we got access to the eigenfrequency spectra and consequently performed an unfolding procedure (For more details, see SI index, section 5). The resulting distributions of NNS of truncated Tannery's pears with $c=0,1,2,5$ are plotted in Fig. \ref{figure55}K. In order to eliminate the effect of artificially-chosen interval on the distribution curve, we adopt the cumulative distribution instead of widely-used histogrammic representation. One can clearly observe that distributions of NNS of all the four cases interpolate between Poisson and WD distributions, indicating that these surfaces are mixing systems. Comparing the distributions of the four cases,
especially in the range of small spacing $s$ which is zoomed in in the inset,
the distribution curve gradually deviates from Poisson distribution and
inclines to WD distribution with the increase of parameter $c$, manifesting a
tendency of increasing chaoticity.

\begin{figure}[ptb]
\centering
\includegraphics[width=8.5cm]{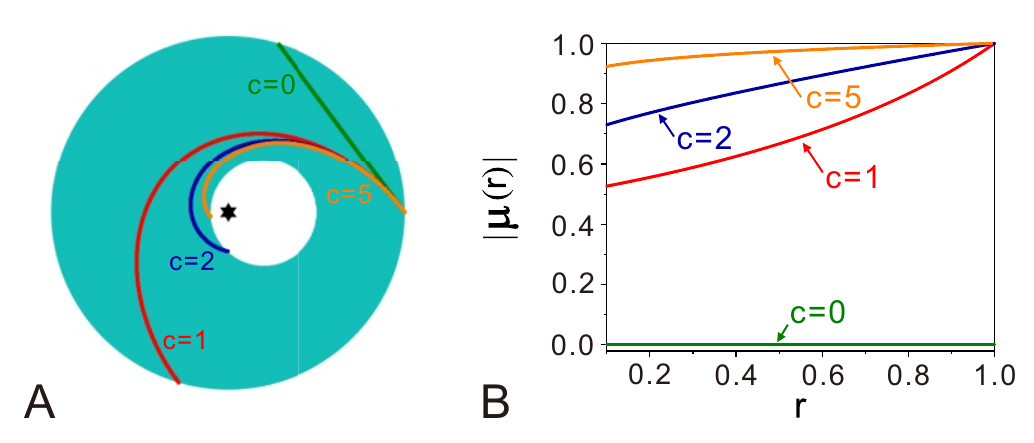}\caption{\textquotedblleft Fictitious
force\textquotedblright\ in projected billiards of Tannery's pears with
different $c$. (A) Comparison of trajectories with same initial conditions,
$\theta=0$ and $p_{\theta}=0.8$, in projected billiards of Tannery's pears
with $c=0$ (green), $1$ (red), $2$ (blue) and $5$ (orange). The star marks the
position of the center of billiard. (b) Comparison of the four surfaces on
$\left\vert \boldsymbol{\mu}(r)\right\vert $, the quantity defined to
characterize the degree of trajectories' deflection.}%
\label{figure7}%
\end{figure}

\section{Discussion}


The above-mentioned three signatures explicitly demonstrate an increasing
tendency of chaoticity with the geometrical parameter $c$. In this section we propose a
qualitative interpretation on this phenomenon. One should note that it is the
unparallel lower boundary on the 3D\ surface/hole in the projected billiard,
or namely, the breaking of symmetry that gives birth to the chaotic dynamics
in the first place \cite{Bunimovich2018}. On this basis, the presence of
curved surface further alters the chaoticity. There are two parallel
approaches to conduct the analysis, either on 3D surfaces or in 2D projected
billards. When considering from the perspective of 3D surfaces, on different
SORs trajectories follow their respective geodesics, the undisturbed natural
paths. Yet the lower unparallel boundaries, which are the reciprocal
projection of the hole in the projected billiards, have different shapes,
leading to different degrees of chaos. However, the various properties of
surfaces, such as the expression of generatrix, curvature, etc, could perplex
the discussion. Whereas, after being projected onto a plane, all the
information relevant to chaotic dynamics is completely and merely embodied in
the distribution of reflective index (since the location and radius of the
hole are fixed in all projected billiards with varying $c$), which simplifies
the problem to a great extent. Fig. \ref{figure7}A compares trajectories
with identical initial conditions in the projected billiards of Tannery's
pears with parameter $c=0,1,2,5$. One can observe visually that with the
presence of nonuniform refractive index, the trajectories deflect towards the
center of the billiard, as if they were subjected to a \textquotedblleft
fictitious force\textquotedblright. More importantly, in the projected
billiard with greater parameter $c$, the trajectories experience more
conspicuous deflections, enhancing the chance of collision on the hole. Put in
another way, a greater initial tangential momentum $p_{\theta}$ is required to
realize the whispering gallery orbits, resulting in a larger area of chaotic
sea and consequently a greater chaoticity of the system. We surmise that the
degree of deflection is proportional to the rate of variation of refractive
index, and hypothetically quantify the \textquotedblleft fictitious
force\textquotedblright\thinspace\ to be
\begin{equation}
\boldsymbol{\mu}=\frac{r}{n(r)}\boldsymbol{\nabla}n(r), \label{miu}%
\end{equation}
with $r$ being the radial coordinate, and $n(r)$ being refractive index
profile. Details about construction of Eq. (\ref{miu}) are revealed in SI index, section 6. Owing to the azimuthal symmetry of SORs, $\boldsymbol{\mu}$ is along the radial direction. Further calculation shows
that the magnitude of $\boldsymbol{\mu}$
is determined uniquely by the metric of the surface (see SI index, section 7). Besides, one may observe
that the term in the absolute value sign is always negative, 
indicating that in the projected billiards of any SORs, the \textquotedblleft fictitious force\textquotedblright%
\thinspace\ always points towards the center of billiard. Fig. \ref{figure7}B illustrates the increasing trend of $\left\vert \boldsymbol{\mu
}(r)\right\vert $ with parameter $c$, which coincides with our assumption. Our interpretation is supported by Fig. \ref{figure3}A, which shows the
dependence of the Lyapunov exponent with parameter $c$ for 3 different positions and diameters of the hole. The difference between these three curves
is a direct consequence of both the pierced hole and presence of refractive
index. Indeed, when the hole is moved away from the center of billiard, or its
size reduces, the billiard is farther from being symmetric, leading to a
higher degree of chaos. In the meantime, areas with higher refractive index
are free from being eliminated and are revealed, further amplifying this difference.

We further inspect the general applicability of quantity $\left\vert
\boldsymbol{\mu}(r)\right\vert $ in SI index, section 7, by exploring some other typical SORs and comparing their $\left\vert
\boldsymbol{\mu}(r)\right\vert $s with the volume of chaotic area in their
Poincar\'{e} SOSs. The results confirm the universality of $\left\vert
\boldsymbol{\mu}(r)\right\vert $. When the studied objects are generalized
from Tannery's pears to general SORs, different SORs are parameterized by
different parameters, and even for SORs which are from a same family, the
relation between their parameters and the degree of chaos might not be as
simple as it is in Tannery's pears: that is where a universal quantity could
play a significant role. Another remarkable advantage of defining $\left\vert
\boldsymbol{\mu}(r)\right\vert $ is that given a SOR, one is able to
approximately estimate its degree of chaos by simply calculating its
$\left\vert \boldsymbol{\mu}(r)\right\vert $, which can be directly obtained
by its metric, instead of investigating more involved signatures of chaos, as
we did earlier.

\section{Conclusion}

In conclusion, we have investigated photonic chaotic dynamics (ray and wave behaviors) on both 3D uniform SORs and in 2D table billiards with nonuniform refractive index. By applying a conformal coordinate transformation, these two systems are proved to be essentially equivalent for both light rays and waves. This complete equivalence enables these two different systems to share interesting phenomena and serve one another to solve in a simpler manner a complex problem, since a geometry can appear easy and analytical in one system, but untraceable in the other. The present proposal, projecting a 3D SOR into a nonuniform table billiard, serves as an innovative pathway to help both solve calculation and simulation problems and provide with a natural interpretation of the role of curvature in 3D chaotic systems, which probably could be done on the 3D SOR  as well but are much simpler and more straightforward to investigate in the latter. Compared with 3D surfaces, a nonuniform 2D table billiard turns out to be a more promising candidate for experimental design and realization, on account of the difficulty in restricting and controlling lights on the former. Reciprocally, the landscape of refractive index in nonuniform table billiards can be complex, while transferring it to its corresponding 3D SOR could greatly reduce the parameters. Our investigation also raises an interesting proposal on the design of refractive index in a nonuniform billiard or cavity to achieve expected features, taking advantage of special trajectories or geometrical properties of its corresponding 3D surface. For instance, it is still an open question for how to transfer an arbitrarily deformed chaotic billiard into integrable, by correlating its refractive index profile with a SOR. More underlying physics and potential applications about this system equivalence remain to be explored.

Although we have taken a typical family of SORs, Tannery's pears, as an example, to study the variation of the degrees of chaos with respect to its control parameter, the model can certainly be generalized to arbitrary SORs. The parameter-dependent property of chaotic dynamics on 3D SORs enlightens an efficient and neat mechanism to control and utilize chaos, and consequently opens up many perspectives, for example, exploration and control of wave chaos in multimode fiber amplifiers to enhance pump absorption efficiency \cite{Legrand2003, Mortensen2007} with nonuniform transverse, or the extensive study of a new type of microcavity \cite{Cho1998, Stone1997, Kim2004, Hentschel2008}. Furthermore, exploration of chaotic dynamics within the context of non-Euclidean geometry may open a new perspective in studies of chaos in cosmology \cite{Bruni2019, Motter2003, Shellard1998}.

\section*{Acknowledgments}

The authors thank Tsampikos Kottos from Wesleyan University for fruitful
discussion on calculation of Lyapunov exponent, and Kun Tang from Bar-Ilan
University for help in COMSOL simulation. C. X. acknowledges the 2019 Israeli
\textquotesingle Sandwich Doctorate Program\textquotesingle\ for international
students funded by the Council for Higher Education at Bar-Ilan University. P.
S. is thankful to the CNRS support under grant PICS-ALAMO. This research is
also supported by The Israel Science Foundation (Grants No. 1871/15, 2074/15
and 2630/20), the United States-Israel Binational Science Foundation NSF/BSF
(Grant No. 2015694), Zhejiang Provincial Natural Science Foundation of China
under Grant No. LD18A040001, the National Natural Science Foundation of China
(NSFC) (grant No. 11674284 and 11974309), and National Key Research and
Development Program of China (No. 2017YFA0304202).

\end{document}